\documentclass[12pt]{article}
\usepackage{amssymb}
\usepackage{epsfig}
\usepackage{float}
\usepackage{graphicx}
\usepackage{slashed}

\newcommand{\be}{\begin{eqnarray}}
 \newcommand{\ee}{\end{eqnarray}}
 \newcommand{\nee}{\nonumber\end{eqnarray}}
 \newcommand{\nn}{\nonumber\\}

  \newcommand{\bc}{\begin{center}}
 \newcommand{\ec}{\end{center}}
  \def\a               {\alpha}
\def\b               {\beta}
\def\m             {\mu}
\def\n              {\nu}

\def\s              {\sigma}
\def\g              {\gamma}

\def\t          {\tau}

\begin{document}
\begin{center}
{\bf  Polarization of the final nucleon
in  quasi-elastic neutrino
scattering and the axial form factor of the nucleon }
\vspace{7mm}

{ \bf Samoil M. Bilenky} \vspace{.1cm}

 {\em  Joint Institute for Nuclear
Research, Dubna, R-141980, Russia;\\}

{\em TRIUMF 4004 Wesbrook Mall, Vancouver BC, V6T 2A3 Canada\\}

\end{center}

\begin{center}
{\bf  Ekaterina Christova}
\vspace{.1cm}

{\em  Institute for  Nuclear Research and Nuclear Energy of BAS,\\ Sofia 1784, Bulgaria\\}
\end{center}
\vspace{.1cm}

\begin{abstract}
Measurements of the polarization of the final proton in elastic
$e-p$ scattering drastically changed our knowledge about  the
electromagnetic form factors of the proton. Here we present
 our results  of the calculation of the polarization of the
final nucleon in charged current quasi-elastic  neutrino nucleon
scattering. Relations which connect the axial form factor with the
polarization, the cross section and the electromagnetic form factors
of the nucleon are derived. Measurements of the polarization of the
nucleon in the high-statistics short baseline neutrino experiments
(or in near detectors of  long baseline experiments) could provide
important information on the axial form factor of the nucleon.
\end{abstract}


\section{Introduction}
Weak and electromagnetic nucleon form factors  are  an important source of information
about the structure of the nucleon. Their study is one of the central issues in high energy physics.

The electromagnetic form factors are determined via investigation of  elastic
 scattering of electrons or muons on proton, deuterium and other nuclei
 (see, for example, reviews \cite{Perdrisat,Arrington}).

Starting from the famous Hofstadter experiments in the 50's and up
to the middle of the 90's,  information about the electromagnetic
form factors  of the proton and the neutron was obtained from
measurements of the differential cross section of unpolarized
electrons on unpolarized nucleons. The electric $G_{E}(Q^{2})$ and
magnetic $G_{M}(Q^{2})$ form factors  of the nucleon were extracted
from these data by the Rosenbluth procedure  based on one-photon
exchange approximation. Until the recoil polarization measurements,
from compilation of the data  it was found:
\begin{enumerate}
  \item The proton form factors satisfy the  approximate scaling relation:
\begin{equation}\label{scaling}
R(Q^{2})=\frac{\mu_{p}G^{p}_{E}(Q^{2})}{G^{p}_{M}(Q^{2})}\simeq 1,
\end{equation}
where  $\mu_{p}$ is the total magnetic moment of the proton (in
nuclear Bohr magnetons),  $Q^2$ is the squared four-momentum
transfer.

\item  At relatively small $ Q^{2}$ ($Q^{2}\leq 6~ GeV^{2}$)
  the  $ Q^{2}$-dependence of the proton form factors and the magnetic form factor of the neutron are described by the dipole formula
 \begin{equation}\label{dipole}
 G^{p}_{M}(Q^{2})\simeq\mu_{p}~G_{D}(Q^{2}),
\quad G^{n}_{M}(Q^{2})\simeq\mu_{n}~G_{D}(Q^{2})
\end{equation}
Here $\mu_{n}$ is the magnetic moment of the neutron and
$$G_{D}(Q^{2})=\frac{1}{(1+\frac{Q^{2}}{M^{2}_{D}})^{2}},$$
where $M^{2}_{D}=0.71~\mathrm{GeV}^{2}$.
\end{enumerate}

In the late 90's
series of experiments on  measurement of the polarization of the recoil protons in elastic scattering of longitudinally
polarized electrons on unpolarized protons started.

In ref. \cite{AkhRek}  it was shown  that measurement of
polarization effects in  elastic $e-p$ scattering
provides a  sensitive way for determination of the electric
form factor of the proton. For  the ratio of the transverse $P_\perp$ and
longitudinal $P_\parallel$  polarizations of the proton it was found
\cite{AkhRek,Arnold}:
\begin{equation}\label{ratio}
\frac{P_\perp}{P_\parallel}=-\,\frac{G^{p}_{E}}
{G^{p}_{M}}\sqrt{\frac{2\varepsilon}{\tau(1+\varepsilon)}},
\end{equation}
where $\tau=Q^{2}/4M^{2}$ ($M$ is the nucleon mass)
and $\varepsilon=[1+2(1+\tau)tan^{2}\theta/2]^{-1}$ ($\theta$ is the scattering angle).

Thus,  measurement of the ratio $P_\perp /P_\parallel$ allows
to determine the ratio of the electric and magnetic form factors in a direct model independent way.

Such measurements were done in  experiments performed in the Jefferson Lab: in  experiments \cite{JLab-1}
in  the $Q^2$ range
from 0.5 to 5.6 $GeV^2$,  in the experiment \cite{JLab-2}
 the  $Q^2$ range was extended up to $Q^2\simeq 8.5$ $GeV^2$.
It was established that  eq.(\ref{scaling}) does not hold and the
ratio $R$ of the electric and magnetic form factors  of the proton
is not a constant but decreases linearly with $Q^{2}$ starting from
$R\simeq 1$ at $Q^2 \simeq 1 ~\mathrm{GeV}^{2}$ and falling down to
$R= 0.28\pm 0.09$ at $Q^{2}=5.6 ~\mathrm{GeV}^{2}$.

These observations  significantly changed  the  theoretical models for the structure of the nucleon.

\vspace{.5cm}

 Direct information about the axial form factor of the nucleon, which characterizes the one-nucleon matrix
 element of the charged weak current, can be
obtained from  measurement of the cross sections of the charged current quasi-elastic (CCQE) neutrino processes:
\begin{equation}\label{numu}
 \nu_{\mu}+n\to \mu^{-} +p
\end{equation}
and
\begin{equation}\label{antinumu}
    \bar \nu_{\mu}+p\to \mu^+ +n,
\end{equation}
which are the  dominant neutrino processes  at relatively small neutrino energies ($ E\leq $ 1 GeV).

Starting with the earlier bubble chamber experiments  many  experiments on the measurements  of the cross sections
of these processes in a wide range of $Q^{2}$ have been done.
 However, the results of these experiments  are not compatible
with each other: from analysis of recent lower $Q^{2}$-data
significantly larger values of the parameter $M_{A}$, which characterizes the $Q^{2}$-dependence of the axial
  form factor, have been obtained.
 Among different reasons for such a
  disagreement nuclear effects are actively discussed.

The axial form factor of the nucleon is of  fundamental importance for the theory.
A  knowledge of the cross sections of the CCQE processes (\ref{numu}) and (\ref{antinumu})
 in a wide range of energies is extremely important for a correct interpretation of
 the  high-precision neutrino oscillation experiments.
 At present, several new experiments
(MINER$\n$A\cite{MINERVA}, T2K\cite{T2K}, ArgoNeuT \cite{ArgoNeuT}) on a   detailed study
 of  CCQE neutrino scattering  are going on.
  In the next Section we will briefly summarize the present day status of  the axial form factor of the nucleon.

Measurement of the polarization of the recoil nucleons in the CCQE processes could be a source
of an important information on the axial form factor of the nucleon. Such measurement, like in the electromagnetic case,
could change our ideas about the
$Q^{2}$-dependence of the axial form factor, about  nuclear effects etc. It is worthwhile and timely to consider  the
possibility for  measurement of  the recoil polarization in modern short baseline neutrino experiments in which
 thousands of neutrino events are detected.

In this paper we shall present the results of the calculations of the recoil polarization of
the nucleon in the CCQE neutrino processes (\ref{numu}) and (\ref{antinumu})
 in the case of the monochromatic neutrino beam  on a free nucleon.

   However, in order to obtain measurable quantities in neutrino (antineutrino) experiments
  one has to average over the neutrino (antineutrino)  spectrum.
  This implies that   to obtain the measurable polarization one needs to average the expressions presented below
  over this spectrum. Note that the numerator and the denominator in the expressions  (\ref{polar}) and
  (\ref{polar1}) must be averaged separately.  Also, in modern neutrino experiments nuclear targets are used.
   Here we do not consider  nuclear effects.

\section{The axial form factor of the nucleon}

The determination of the axial form factor of the nucleon is a very challenging experimental problem
 due to the fact that in neutrino experiments  nuclear targets (C, Fe, etc.) are used, the neutrino beams
are not monochromatic, they are normalized in different ways etc.

In analogy with the electromagnetic form factors the axial form
factor  is usually parameterized by the dipole formula:
\begin{equation}\label{dipoleGA}
G_{A}(Q^{2})=\frac{g_{A}}{(1+\frac{Q^{2}}{M^{2}_{A}})^{2}}.
\end{equation}
Here  $g_{A}=1.2701\pm 0.0025$ \cite{gA}  is the axial constant,
determined from the neutron $\beta$-decay data
 and $M_{A}$ is a parameter ( the "axial mass").

 The values of the parameter $M_{A}$ determined from the data of different experiments, under
 the assumption that neutrinos interact with a quasi-free nucleon in a nuclei and other nucleons are spectators
 ( impulse approximation), are quite different.

From analysis of the data on  measurements of the cross section of the process $\nu_{\mu}+n\to \mu^{-} +p$
 on deuterium target and of the process  $\bar\nu_{\mu}+p\to \mu^{+} +n$
on  proton target it was found  \cite{Bodek}:
\begin{equation}\label{Bodek}
 M_{A}=1.016\pm 0.026~\mathrm{GeV}.
\end{equation}

The value  of the parameter
 $M_{A}$ obtained from the data of the NOMAD experiment (carbon target)
 \cite{Nomad} is in agreement with (\ref{Bodek}):
 \begin{equation}\label{Nomad}
 M_{A}=1.05\pm 0.02 \pm 0.06~\mathrm{GeV}.
\end{equation}
Let us note that the  value (\ref{Nomad}) was found from the total
cross section averaged over the neutrino spectrum.  In the same
experiment, but  from the $Q^{2}$-distribution a value  of the
parameter $M_{A}$  \cite{Nomad}:  $ M_{A}=1.07\pm 0.06 \pm
0.07~\mathrm{GeV}$ was extracted, which is compatible with
(\ref{Nomad}), but has larger statistical and systematic errors.

However, from  fit  of the data of
more recent experiments larger average values of the parameter
$M_{A}$ (with larger errors) were obtained.

From the data of the MINOS experiment  (iron target) it was found~\cite{Minos}:
\begin{equation}\label{Minos}
 M_{A}=1.26 _{-0.10}^{+0.12} {}_{-0.12}^{+0.08}~\mathrm{GeV}.
\end{equation}

 In the K2K experiment ($H_{2}O$ target) it was obtained~\cite{K2K}:
\begin{equation}\label{K2K}
 M_{A}=1.20\pm 0.12~\mathrm{GeV}.
\end{equation}

 From the data of the MiniBooNE experiment (carbon target) it was
found \cite{Miniboone1}:
\begin{equation}\label{Miniboone1}
 M_{A}=1.23\pm 0.20~\mathrm{GeV}.
\end{equation}

From the analysis of data of the later high-statistic
experiment\cite{Miniboone} ($1.4\cdot 10^{5}$ events) it was
 inferred:
\begin{equation}\label{Miniboone}
 M_{A}=1.35\pm 0.17~\mathrm{GeV}.
\end{equation}

There could be many different reasons for the disagreement of the average values of  $M_{A}$
obtained from  the data of the different experiments.
It could be a problem of systematics and normalization (see~\cite{Ankowski}).
Target nuclei in the different experiments are different.
 The difference of the values of  $M_{A}$
 could be due to such nuclei effects as interaction of neutrinos with
correlated pairs of nucleons (see \cite{Martini1, Martini2}).
Experiments on the study of CCQE neutrino processes were done in different ranges of $Q^{2}$.
The difference between the different values of $M_{A}$ could be a signature that the dipole parametrization
(\ref{dipole}) may not be the correct parametrization of the axial form factor in the whole region of $Q^{2}$ studied
(like in the case of the electromagnetic form factors).

A measurement of the polarization of the recoil protons produced in the CCQE neutrino
process $\nu_{\mu}+n\to \mu^{-}+p$ could change the situation with axial form factor $G_{A}(Q^{2})$.
Taking into account that in short baseline neutrino experiments thousands of neutrino events are observed
 it is worthwhile to consider a possibility of measuring of the polarization of the protons by the observation
 of left-right asymmetry in the scattering of the recoil protons in a neutrino detector.

 In the next section we will present our results of the calculation
  of the polarization of final nucleon   in the CCQE neutrino processes.

\section{Polarization of the final nucleons in CCQE processes}

Here we shall present the results of the calculations of the polarization of final nucleons in the CCQE neutrino processes
 (\ref{numu}) and (\ref{antinumu}).

Process (\ref{numu}) is a charged current process   and
its matrix element is characterized by the four weak form factors of the nucleon:
\begin{eqnarray}\label{matelem}
\hspace*{-1cm}\langle f|~(S-1)~|i\rangle
&=&-i\,\frac{G_{F}\cos\theta_c}{\sqrt 2}\,N_{k}N_{k'}~
\bar u(k')\gamma^{\alpha}(1-\gamma_{5})\,u(k). ~_{p}\langle p'|~J^{(1+i2)}_{\alpha}~|p\rangle_{n} \nonumber\\
 &&\times (2\pi )^4\,\delta (k+p-k'-p').
\end{eqnarray}
Here $G_F$ is the Fermi  constant, $\theta_c$ is the Cabbibo mixing
angle. The hadronic matrix element $_{p}\langle p'|~J^{(1+i2)}_{\alpha}~|p\rangle_{n}$ is:
\begin{equation}\label{matelem1}
_{p}\langle p'|~J^{(1+i2)}_{\alpha}~|p\rangle_{n}=N_{p}N_{p'}\bar u(p')(V_{\alpha}-A_{\alpha})u(p),
\end{equation}
where
\begin{equation}\label{matelem2}
\hspace*{-1cm}V_{\alpha}=\gamma_{\alpha}F^{CC}_{1}(Q^{2})+\frac{i}{2M}
\sigma_{\alpha\beta}q^{\beta} F^{CC}_{2}(Q^{2}),\quad
A_{\alpha}=\gamma_{\alpha}\gamma
_{5}G_{A}(Q^{2})+q_{\alpha}\gamma_{5} G_{P}(Q^{2}),
\end{equation}
 $F_{1,2}^{CC}$,  $G_{A}$ and $G_{P}$ are the CC weak vector,
axial and pseudoscalar form factors, respectively,
 $k$ and $p$ ($k'$ and $p'$ ) are the
initial neutrino and neutron (final muon and proton) momenta, $N_{p}=\frac{1}{(2\pi)^{3/2}}\sqrt{2p_{0}}$
is the standard normalization factor, $q=p'-p=k-k'$ is momentum transfer, $Q^{2}=-q^{2}$.

 Under  isotopic $SU(2)$ transformations the weak  charged  current $J^{1+i2}_{\alpha}$ is transformed as
  the "plus component" of the  conserved isovector current.
Taking into account that the third component of this isovector is the isovector part of the electromagnetic current
-- the hypothesis for conservation of the vector current (CVC),
 from isotopic $SU(2)$ invariance for the weak vector form factors  we obtain:
\begin{equation}\label{CVC1}
F^{CC}_{1,2}(Q^{2})=F^{p}_{1,2}(Q^{2})-F^{n}_{1,2}(Q^{2}),
\end{equation}
where $F^{p}_{1,2}(Q^{2})$ and $F^{n}_{1,2}(Q^{2})$ are the
Dirac and Pauli electromagnetic form factors of the proton and the neutron.
These form factors  are known at present in a wide region of $Q^{2}$ ( see, for example, the review \cite{Perdrisat,Arrington}).

From the hypothesis for partial conservation of the axial current (PCAC)
 it follows that the contribution of the pseudoscalar form factor  $G_{P}(Q^{2})$ to the matrix element (\ref{matelem2})
 can be neglected.  Thus, from  study of the CCQE process
  (\ref{numu}) an information about the axial form factor $G_{A}(Q^{2})$ can be obtained.

 The matrix elements of the processes (\ref{numu}) and
(\ref{antinumu}) are characterized by the same form factors. In fact
from charge symmetry we have:
\begin{equation}\label{chargesym}
_{p}\langle p'|~J^{(1+i2)}_{\alpha}~|p\rangle_{n}=_{n}\langle p'|~J^{(1-i2)}_{\alpha}~|p\rangle_{p}.
\end{equation}

 The polarization 4-vector  of the final proton  in process
(\ref{numu}) is given by the expression:
\begin{equation}\label{polar}
\xi^{\rho}=\frac{\mathrm{Tr}\left[\gamma^{\rho}\gamma_{5}~\rho_{f}\right]}
{\mathrm{Tr}\left[\rho_{f}\right]},
\end{equation}
where $\rho_{f}$ is  the final spin density matrix. Using the
relation
\begin{equation}\label{polar1}
\Lambda(p')\gamma^{\rho}\gamma_{5}\Lambda(p')=2M\left(g^{\rho\sigma}-
\frac{p'^{\rho}p'^{\sigma}}{M^{2}}\right)\Lambda(p')\gamma_{\sigma}\gamma_{5}
\end{equation}
and performing integration over the momenta of the final lepton and
nucleon, for  the polarization 4-vector of the final nucleon we
have:
\be
\xi_\a=\left( g_{\a\b}-\frac{{p'}_\a
p'_\b}{M^2}\right)\,\frac{Tr\left[{\cal N}\Lambda (p)\bar {\cal N}
\Lambda (p')\g^\b \g^5\right]} {Tr\left[{\cal N} \Lambda (p)\bar
{\cal N} \Lambda (p')\right]}.\label{xi}
 \ee
 Here
 \be
  {\cal N}&=&\,\bar u(k')\g_\a (1-\g_5)u(k)\,\left(V^\a - A^\a \right),
\ee
$\Lambda(p)=p\!\!/ +M$. In eq. (\ref{xi}) the projection
operator $(g_{\a\b}-p'_{\a}p'_{\b}/M^{2}) $ guarantees the condition
$(\xi\cdot p')=0$.

The vector $\xi^\a$ can be decomposed along the following three
independent 4-vectors $Q_i^\a$ orthogonal to $p'^\a$:
\be
 Q^\a_+
=k_+^\a -\frac{(p' k_+)}{M^2}\,{p'}^\a , \quad
Q^\a_-={k^\a_-}-\frac{(p'k_-)}{M^2}\,{p'}^\a,\quad Q^\a_p
={p}^\a-\frac{(p'p)}{M^2}\,{p'}^\a\label{Q} \ee where \be
k_+=(k+k'),\qquad k_-=(k-k')=q.
\ee

After  standard calculations  we obtain:
\be
 \xi^\a
&=&\frac{M}{(kp)\,J_0}\,\left[\, Q^\a_+\,P_++
Q^\a_{-}\,P_-+ Q^{\a}_p\,P_p\,\right]\label{final}\\
 P_+&=& \left[y\,G_M^{CC}+(2-y)G_A\right] \,G_E^{CC}\label{P+}\\
P_-&=& -\, G_A\left[y\,G_M^{CC}+(2-y)G_A\right]+
\,F_2^{CC}\,\left[\,(2-y)\,\tau G_M^{CC}+y\,(1+\tau)\,G_A\right]\label{P-}\nn\\
P_p&=&\,\,\frac{F_2^{CC}}{y}\,\left[2y(2-y)\,\tau\,G_M^{CC}+\left[2\t
[1+(1-y)^2]+y^2\right]G_A\right].\label{Pp} \ee Here \be
  G_E^{CC}&=&\,F^{CC}_{1}-\tau F^{CC}_{2}, \qquad G_M^{CC}=F^{CC}_{1}+F^{CC}_{2}\nn
 J_0&=&\frac{Tr\left[{\cal N} \Lambda (p)\bar
{\cal N} \Lambda (p')\right]}{8^2(kp)^2},\quad y=\frac{(pq)}{(pk)},\quad \tau = \frac{Q^2}{4M^2}.
  \ee
  From (\ref{CVC1}) it follows:
\begin{equation}\label{polar10}
 G^{CC}_{M}=G^{p}_{M}- G^{n}_{M} ,\quad  G^{CC}_{E}=G^{p}_{E}- G^{n}_{E},
\end{equation}
where $G^{p,n}_{M}$ and $G^{p,n}_{E}$ are the magnetic and charge form factors of proton  and neutron.

\vspace{.5cm}

 From (\ref{final}) one  can easily find the  polarization vector  of the proton in the laboratory frame. We have:
 \begin{equation}\label{xilab}
\hspace*{-1cm}\vec{\xi}=\frac{1}{J_0\,E}\left\{(\vec{k}+\vec{k'})P_{+}+
\vec{q}~\left[-\frac{E+E'}{M}~P_{+}+(1+\frac{E-E'}{M})~\left(P_{-}-P_p\right)\right]\right\}.
\end{equation}
Here
$E$ and $E'$ are the energies of the neutrino and the final muon:
 \begin{equation}\label{trans1}
E'=\frac{E}{1+(2E/M)\sin^2(\theta/2)},\qquad y=\frac{E-E'}{E},
\end{equation}
$\theta$ is the angle between the vectors $\vec{k}$ and $\vec{k'}$.

  The  polarization vector lays in the scattering
  plane.\footnote{It is obvious that the component orthogonal to the scattering plane disappears due to $T$-invariance.}
  For the longitudinal $\xi_\|$ and transverse $\xi_\bot$ components
  of the polarization we have:
 \begin{equation}\label{polarLab}
\vec{\xi}=\xi_{\bot}\vec{e}_{\bot}+\xi_{\|}\vec{e}_{\|},
\end{equation}
where $\vec{e}_{\bot}$ and $\vec{e}_{\|}$ are two orthogonal unit vectors in the scattering plane:
\begin{equation}\label{vectors}
\vec{e}_{\|}=\frac{\vec{p'}}{|\vec{p'}|}=\frac{\vec{q}}{|\vec{q}|},\qquad
\vec{e}_{\bot}=\vec{e}_{\|}\times \vec{n},\qquad  \vec{n}=\frac{\vec{k}\times\vec{k'}}{|\vec{k}\times\vec{k'}|}.
\end{equation}

From (\ref{xilab}), (\ref{polarLab}) and (\ref{vectors}) we obtain:
\be
s_{\bot}=\xi_{\bot}
=\left(\frac{1}{J_0}\right)\frac{-2\,E'\sin\theta}{\vert\vec q\vert}
~\left[G_{A}(2-y)+G^{CC}_{M}y\right]G^{CC}_{E}.\label{trans} \ee \be
s_{\|}&=&\frac{M}{p'_{0}}\,\xi_{\|}=\nn &=& -\,\frac{1}{J_0} \,
\frac{q_0}{\vert \vec q\vert}\,
\left[G_{A}(2-y)+G^{CC}_{M}y\right]\left[G^{CC}_{M}(2-y)+G_{A}(y+
\frac{2M}{E})\right].\label{long} \ee Here $s_{\|}$ and $s_{\bot}$
are the longitudinal and transverse components of the polarization
vector in the rest frame of the recoil nucleon: \be s^\a =
(0;s_{\|},s_{\bot}) \ee $M/p'_0$ is the Lorentz boost along $\vec
p^{\,\,'}$.

  Let us note that at $G_A=0$, using the kinematic relations
\begin{equation}\label{trans1}
|\vec{q}|=Ey\sqrt{\frac{1+\tau}{\tau}},\qquad \frac{q_0}{\vert \vec q\vert}=\sqrt{\frac{\tau}{1+\tau}},
\end{equation}
one can show that eqs. (\ref{trans}) and (\ref{long})
coincide with the well known expressions for the transverse and
longitudinal polarizations of the recoil protons in elastic
scattering of longitudinally polarized leptons on unpolarized
protons (see, for example, \cite{Perdrisat}).

Taking into account that the hadronic part of the processes
$ \nu_{\mu}+n\to \mu^- +p$ and $\bar\nu_{\mu}+p\to \mu^+ +n$ are the same, we easily obtain the
 polarizations of the final nucleons for both processes:
\be
(J_0\,s_{\bot})^{\n ,\,\bar\n} =
\frac{-\,2E'\,\sin\theta}{\vert\vec q\vert}\left[\pm
y\,G_M^{CC}+(2-y)G_A\right]\,G_E^{CC}\label{trans2}
 \ee
 and
 \be
\left(J_0\,s_{\|}\right)^{\n ,\,\bar\n}=-\, \frac{q_0}{\vert \vec
q\vert}\,\left[ \pm y\,G_M^{CC}+(2-y)\,G_A\right]
 \left[(2-y)\,G_M^{CC}\pm \left(y+ \frac{2M}{E}\right)\,G_A\right].\nn\label{long2}
\ee
 Here and further the upper (lower) sign corresponds to
neutrino (antineutrino) scattering.

The quantity  $J_0^{\n ,\,\bar\n}$ is determined from the differential cross section:
\be
J_0^{\n ,\bar\n}=\frac{d\s^{\n ,\,\bar\n}}{dQ^2}\cdot\,\frac{4\pi}{G_F^2\,\cos^2\theta_c}.
\ee
In terms of the form factors it is given by the expression:
 \be
J_0^{\n ,\,\bar\n}
 &=&2(1-y)\left[G_A^2+\frac{\tau (G_M^{CC})^2+(G_E^{CC})^2}{1+\tau}\right]
 +\frac{My}{E}\,\left[\,G_A^2-\frac{\tau (G_M^{CC})^2+(G_E^{CC})^2}{1+\tau}\right]\nn
 &&+y^2\,(G_M^{CC}\mp G_A)^2\pm 4y\,G_M^{CC}\,G_A.
 \ee


\section{Comments }

 -- From eq.(\ref{trans2}) we obtain a rather simple expression for $G_A$:
\be
G_A=\frac{-1}{2-y}\,\left\{\frac{ M \sqrt {\tau
(1+\tau)}}{E'\sin\theta}\, \frac{(J_0 . s_\bot )^{\n ,\bar
\n}}{G_E^{CC}}\,\pm y\,G_M^{CC}\right\}.
\ee

\noindent
--  Note, that the electric form factor does not enter eq.(\ref{long2}). Thus
the axial form factor $G_A$ is determined only by the
cross section, the longitudinal polarization $\xi_\|$
 and the best known magnetic form factors
 of the proton and neutron.

\noindent
--  If the neutrino detector is in a magnetic field, then  both the transverse and longitudinal
  polarizations could be measured (like in the case of elastic $e-p$ scattering). For their ratio we have:
\be
\left(\frac{s_\|}{s_\perp}\right)^{\n
,\bar\n}&=&\frac{q_0}{2E'\sin\theta}\,\, \frac{
\left[(2-y)\,G_M^{CC}\pm \,G_A(y + 2M/E)\right]}{G_E^{CC}}.
\ee
 Then
for the axial form factor we obtain:
 \be
\,G_A=\pm\,\frac{E+E'}{E-E'+2m}\left[\frac{2EE'\sin\theta}{E^2-E'^2}\,G_E^{CC}\,
\left(\frac{s_\|}{s_\perp}\right)^{\n ,\bar\n}\,-{G_M^{CC}}\right].
\ee

\noindent
-- Finally, let us notice the relations:
 \be
 (J_0\,s_\perp)^{\n}+
(J_0\,s_\perp)^{\bar\n} = \frac{-\,4\,E'\,\sin\theta}{\vert\vec
q\vert}\,(2-y)\,G_A\,G_E^{CC}
\ee
 \be
(J_0\,s_\|)^{\n}+(J_0\,s_\|)^{\bar\n} =\frac{-\,4\,q_0}{\vert \vec
q\vert}\,
\,G_A\,G_M^{CC}\left\{\left[1+(1-y)^2\right]+\frac{My}{E}\right\}.
\ee

\section{Numerical results}

 Here we present a numerical study of the sensitivity of the discussed recoil nucleon
polarization to the  different choices of the axial mass
$M_A$.

  We  use  the following commonly used
 parameterizations for the  form factors, summarized in \cite{Perdrisat}:
 \be
 G_D&=&\frac{1}{\left(1+\frac{Q^2}{M_V^2}\right)^2},\qquad M_V^2=0.71\nn
 G_{M,p}&=&\m_p\,G_D,\qquad  G_{M,n}=\m_n\,G_D\nn
G_{E,p}&=&(1.06-0.14\,Q^2)\,G_D\nn
G_{E,n}&=&-a\,\frac{\m_n\t}{1+b\t}\,G_D,\qquad a=1.25,\quad b=18.3 \label{parms}
\ee
 where $\m_p=2.79$ and  $\m_n=-1.91$ are the magnetic moments of the proton and neutron.
 We  calculate the effect of the different axial form factors on
 the longitudinal and transverse polarizations, considering the following values of  $M_A$:\\
\be
&&1) \,\,M_A=1.016\, - \,full \,line\nn
&&2)  \,\,M_A=1.20\,  -\, dashed\, line\nn
&&3)  \,\,M_A=1.35\, -\, dotted \,line\label{MA}
\ee
We examined the polarizations at fixed neutrino energies
 as functions of $Q^2$  in the energy range  $Q^2_{min}\leq Q^2\leq Q^2_{max}$. Here
$Q^2_{min}$ and $Q^2_{max}$ are fixed by the condition $0\geq\cos\theta
\leq1$ and  the scattering angle  $\theta$ is determined via
(\ref{trans1}). We have:
 \be
\cos\theta&=&1-\frac{MQ^2}{E(2ME-Q^2)}\nn
  Q^2_{min}&=&0,\qquad Q^2_{max}=\frac{2ME^2}{M+E}\nn
  \sin\theta&=&\frac{MQ^2}{E(2ME-Q^2)}\,\sqrt{\frac{4E^2}{Q^2}-\frac{2E}{M}-1}.
  \ee

\begin{figure}[htb]
 \centerline{ \epsfig{file=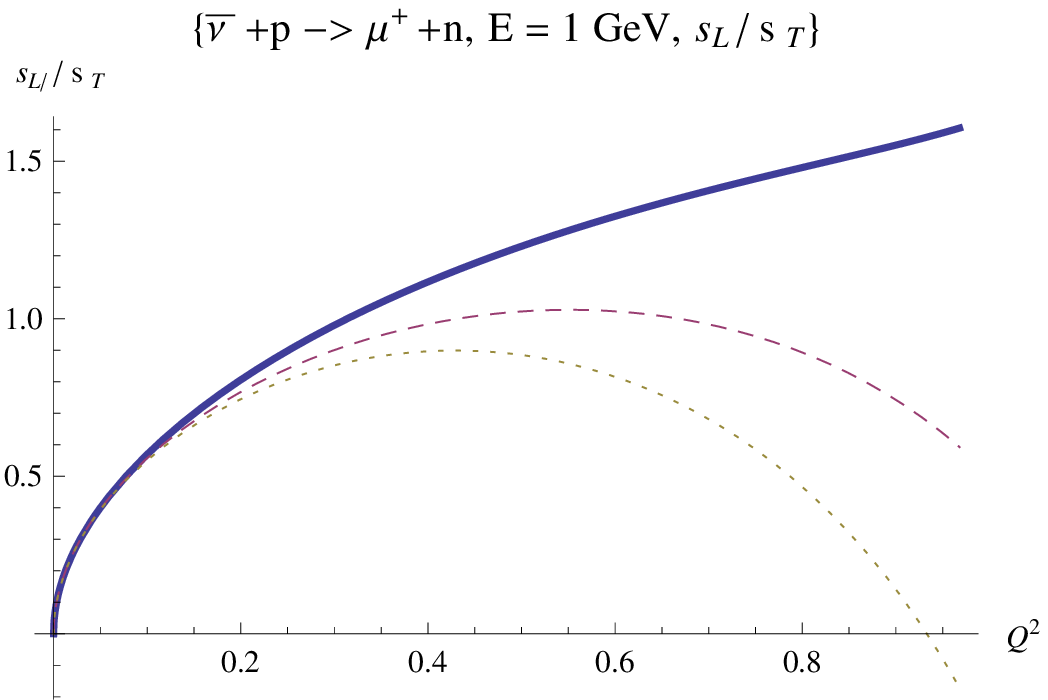,width=60mm}\hspace{7mm} \epsfig{file=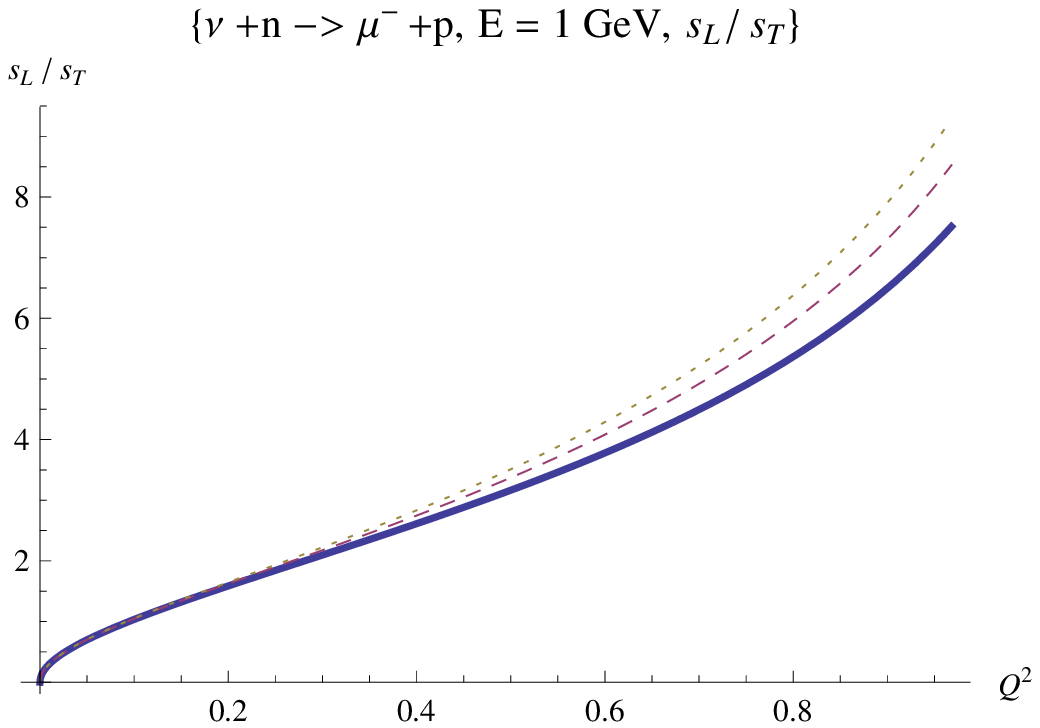,width=60mm}}
  \vspace{7mm}

\centerline{\epsfig{file=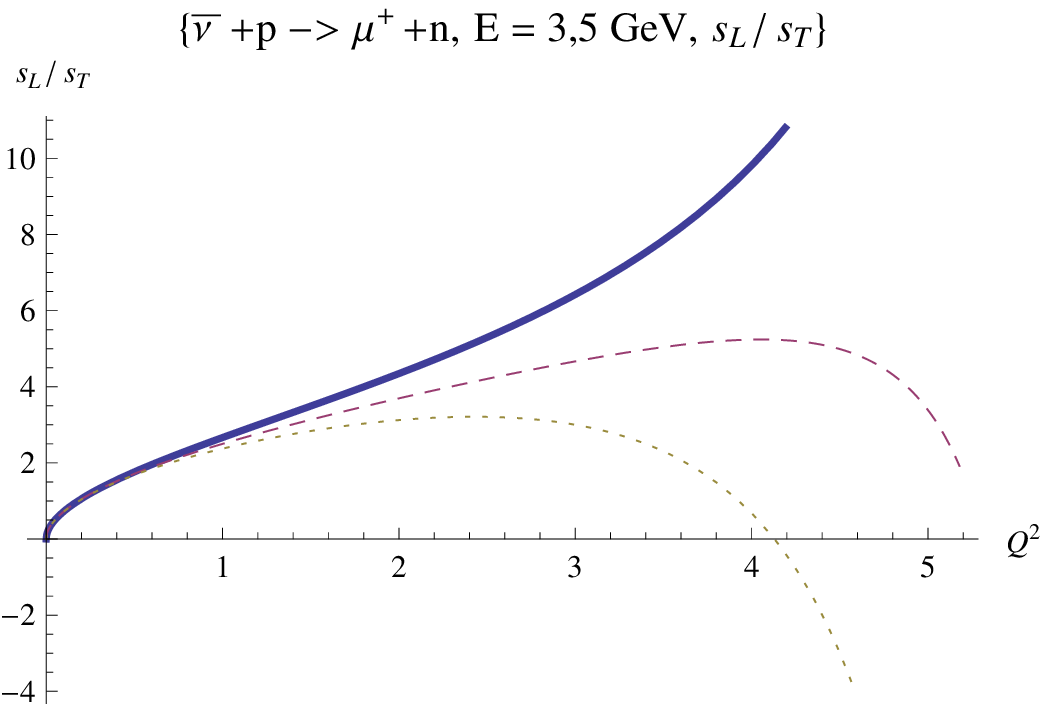,width=60mm}\hspace{7mm} \epsfig{file=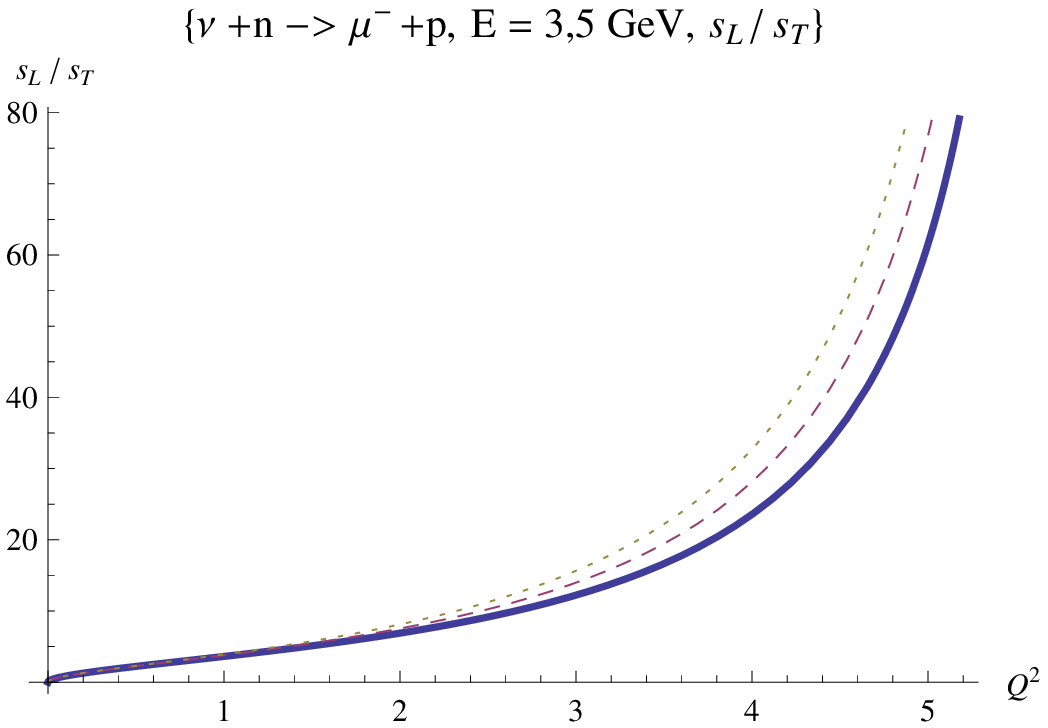,width=60mm}}
 \vspace{7mm}

\centerline{\epsfig{file=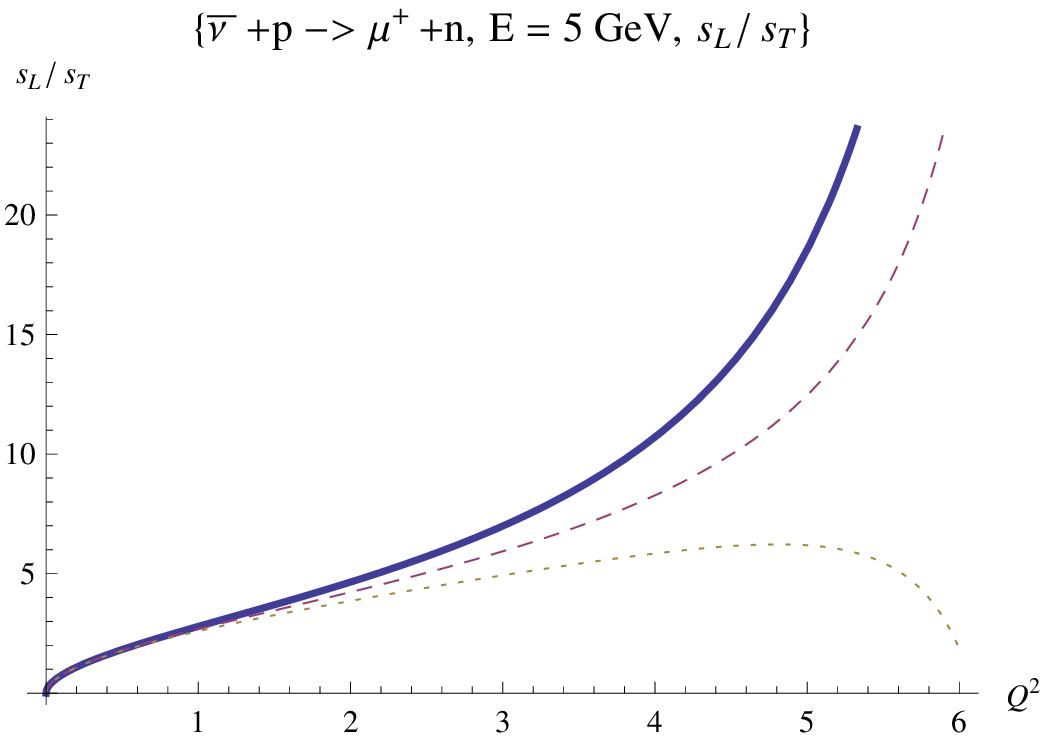,width=65mm}\hspace{7mm} \epsfig{file=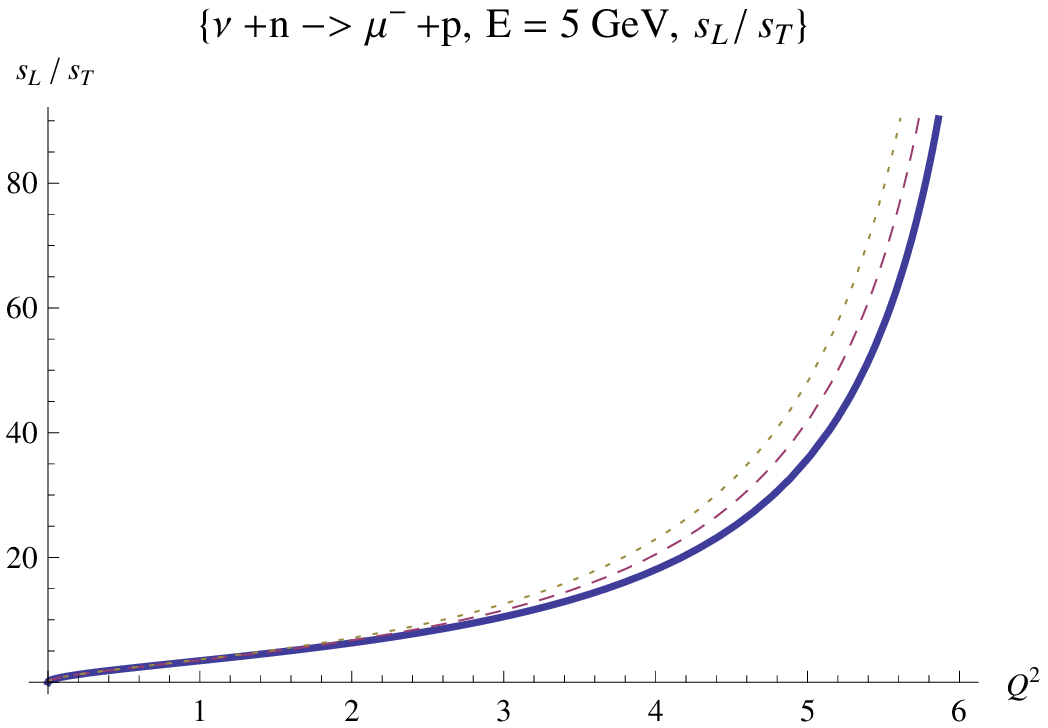,width=65mm}} \caption{The
dependence of $s_L/s_T$ on the values of $M_A$ [see(\ref{MA})]
for $\bar\nu_\mu+p\to \mu^++n$ (\textit{left}) and for $\nu_\mu+n\to \mu^-+p$
 (\textit{right})  at E=1 (\textit{up}),  3,5
(\textit{middle}) and 5 (\textit{down}) GeV.}\label{r}
\end{figure}

\begin{figure}[htb]
 \centerline{ \epsfig{file=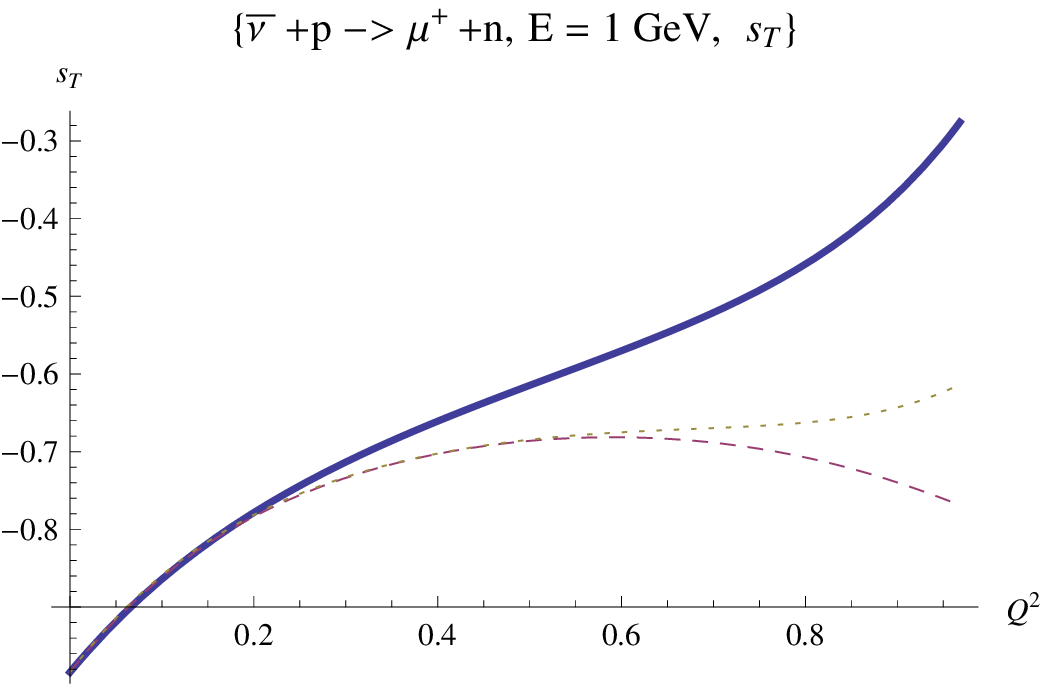,width=60mm}\hspace{7mm} \epsfig{file=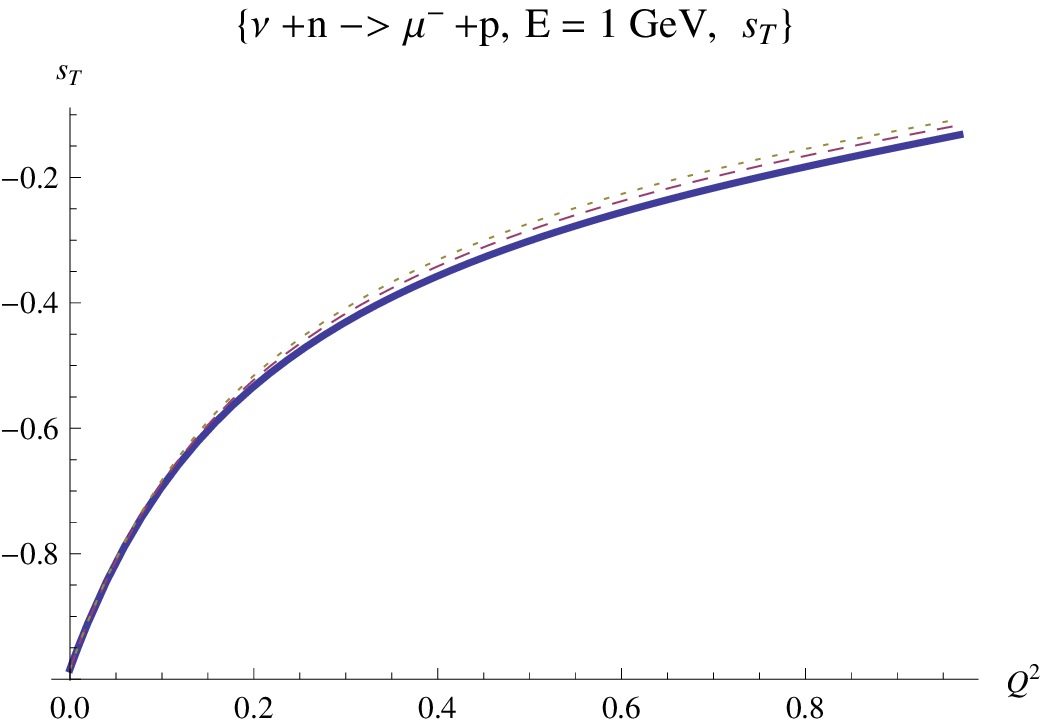,width=60mm}}
  \vspace{7mm}

\centerline{\epsfig{file=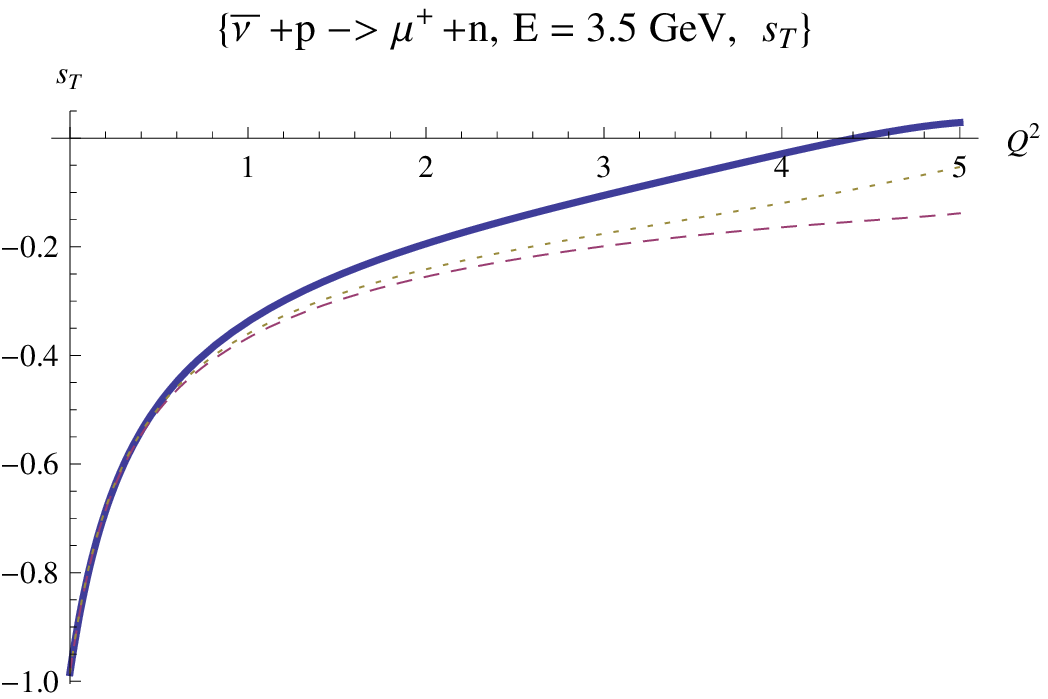,width=60mm}\hspace{7mm} \epsfig{file=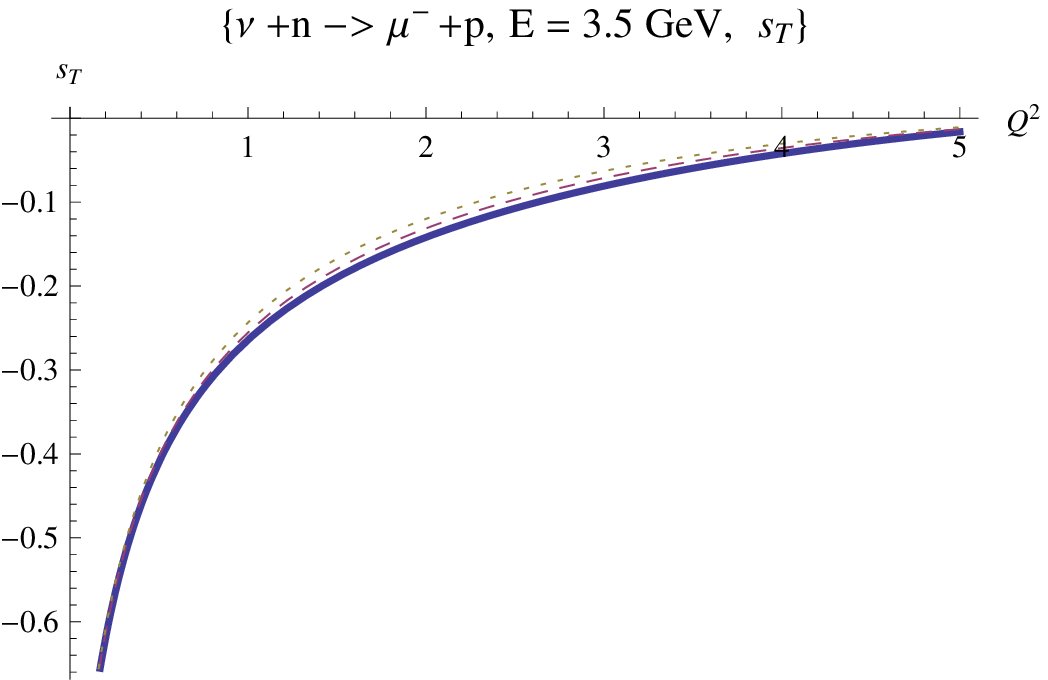,width=60mm}}
 \vspace{7mm}

\centerline{\epsfig{file=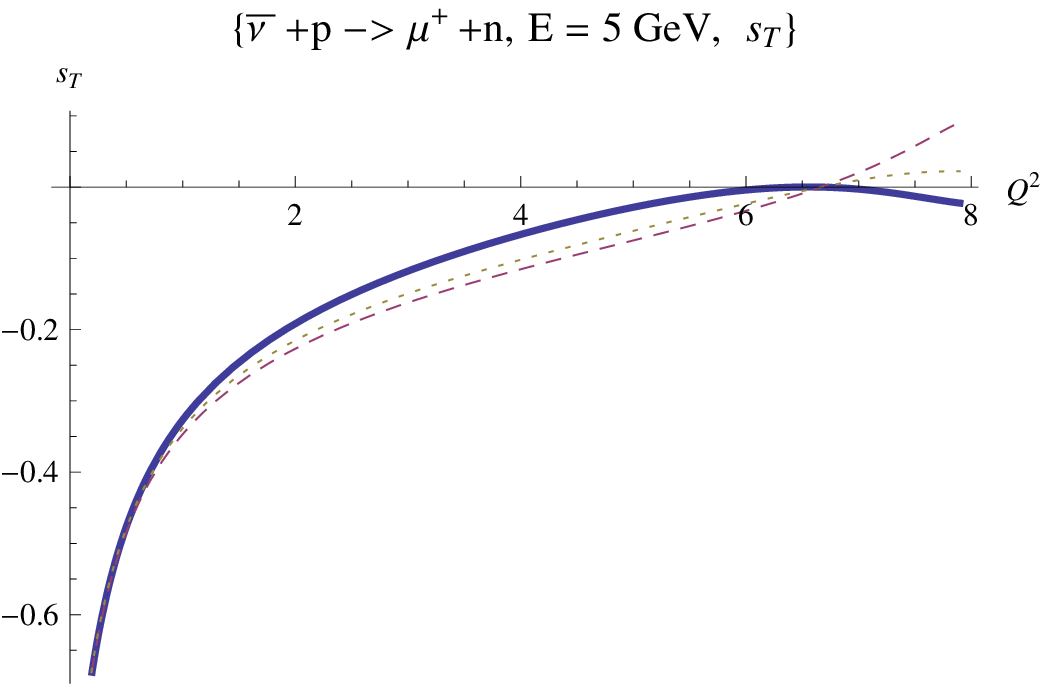,width=65mm}\hspace{7mm} \epsfig{file=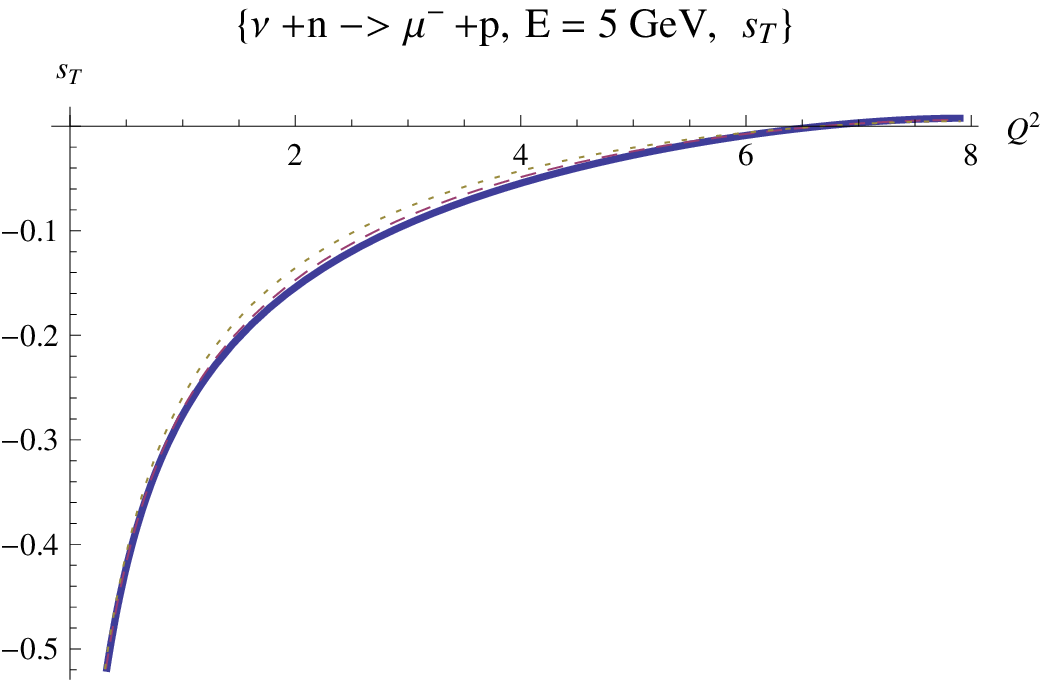,width=65mm}} \caption{The
dependence of the transverse polarization $s_T$ on the  values of $M_A$ [see(\ref{MA})]
for $\bar\nu_\mu+p\to \mu^++n$ (\textit{left}) and for $\nu_\mu+n\to \mu^-+p$
 (\textit{right})  at E=1 (\textit{up}),  3,5
(\textit{middle}) and 5 (\textit{down}) GeV.}\label{sT}
\end{figure}

On Figs.(\ref{r}) and (\ref{sT}) we show the dependence of $s_\|/s_\perp$ and  $s_\perp$
on the choice of $M_A$ for the two considered processes: $\bar\n +p\to \m^+ + n$ (left) and
$\n +n\to \m^- + p$ (right).

We found that the polarization of the final proton in $\nu_{\mu}+n\to \mu^{-}+p$ practically does not depend
on the value of $M_A$. However, polarization of the final neutron in $\bar \nu_{\mu}+p\to \mu^{+}+n$
 is rather sensitive to the value of the axial mass.
 It is most clearly pronounced for the longitudinal polarization $s_\|$ and, respectively, for the ratio
$s_\|/s_\perp$, shown on Fig.(\ref{r}). Note that an advantage of  $s_\|/s_\perp$ is that
many of the systematic uncertainties and radiative corrections cancel, however a magnetic field should be applied
to the detector in order to measure $s_\|$.
 This  sensitivity is exhibited  in the whole $Q^2$-range. As higher $Q^2$ are accessed through higher
 neutrino energies (and also the measured quantities are averaged over the neutrino spectra),
we have presented the  polarizations  for three values of the neutrino energies: E=1, 3.5 and 5 GeV.
  The transverse polarization in $ \nu_{\mu}+p\to \mu^{+}+n$  shows  sensitivity to $M_A$ for
  low neutrino energies. but not so
 dramatically pronounced for higher energies  (see Figs.(\ref{sT})).

\begin{figure}[htb]
 \centerline{ \epsfig{file=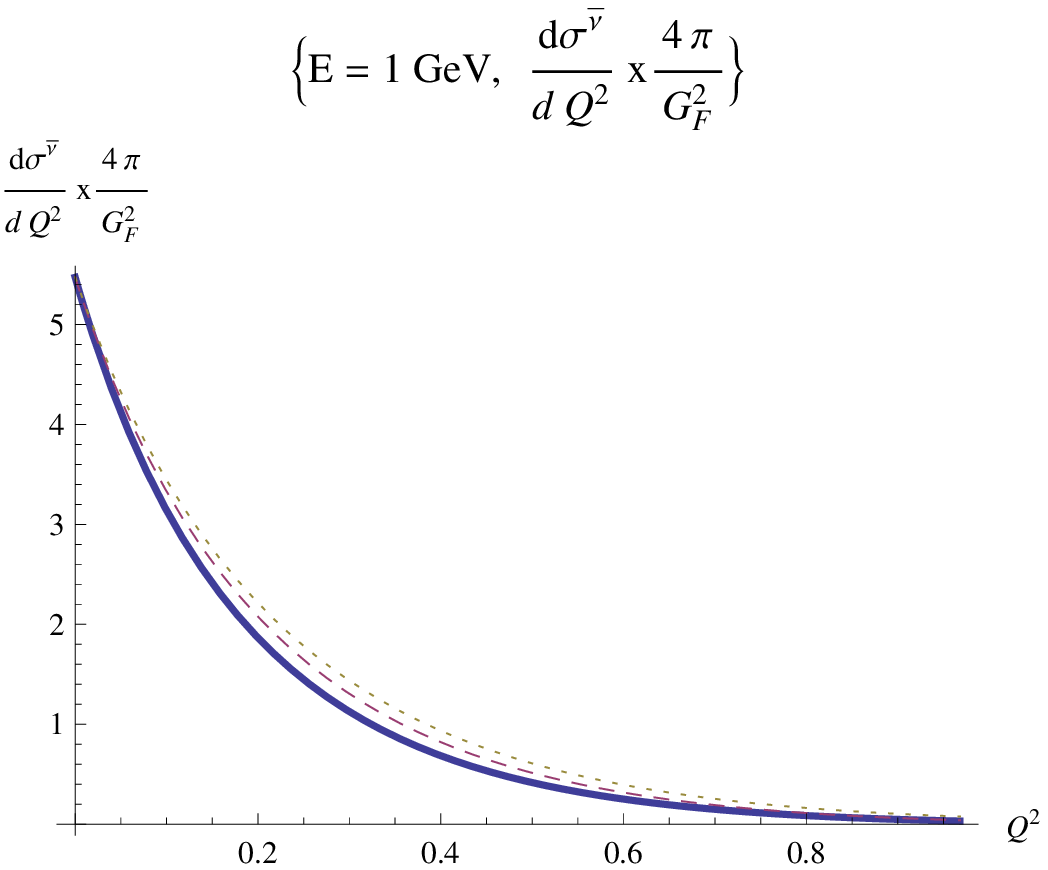,width=65mm}\hspace{0.5cm} \epsfig{file=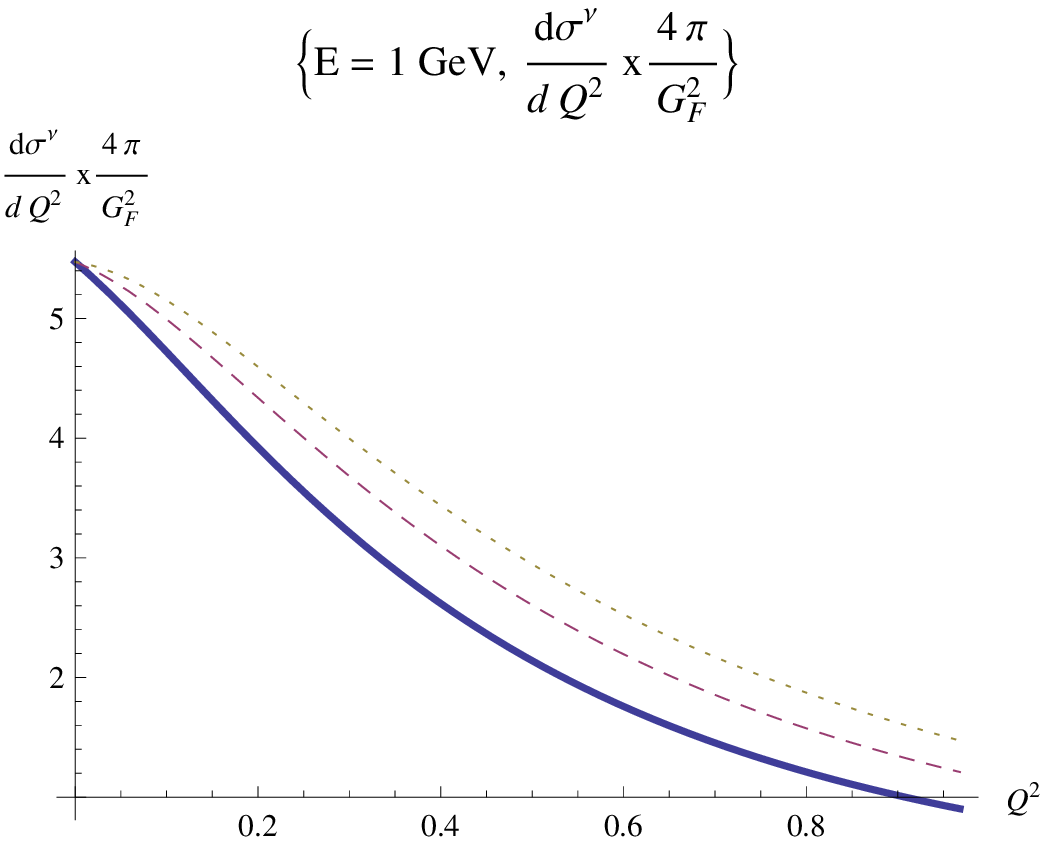,width=65mm}}
  \vspace{3mm}

  \centerline{\epsfig{file=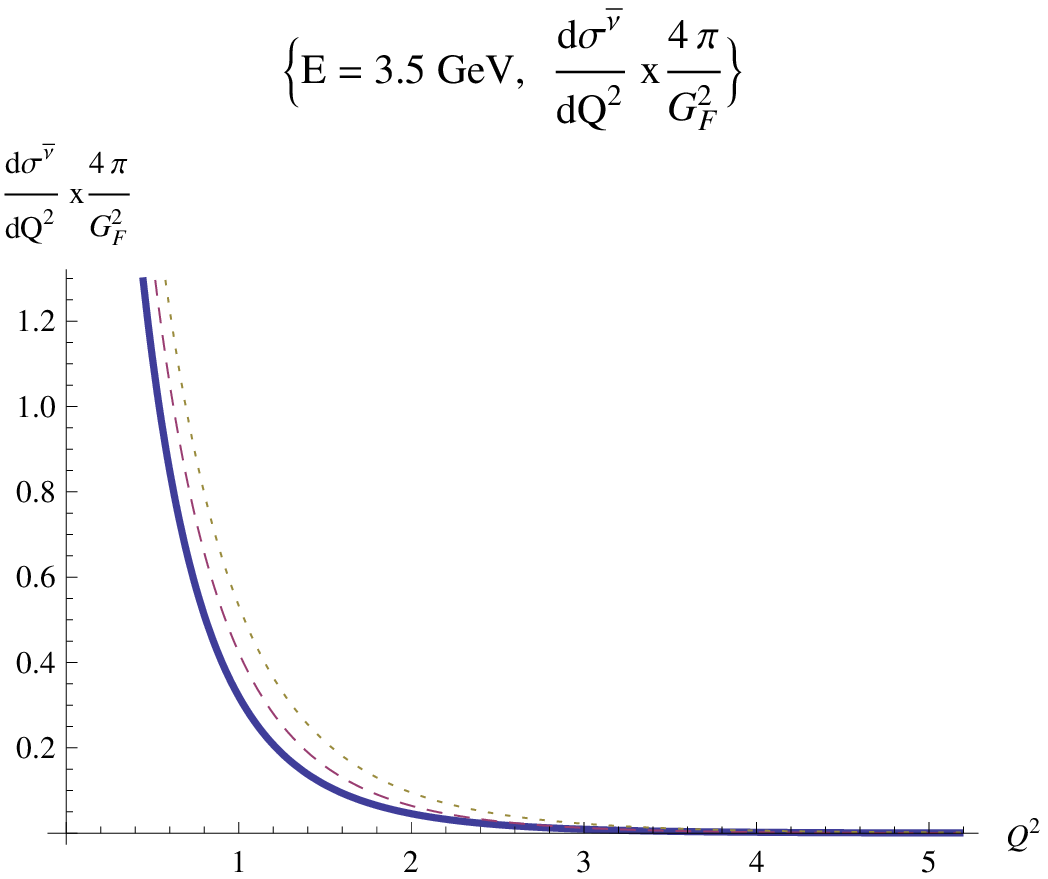,width=65mm}\hspace{0.5cm} \epsfig{file=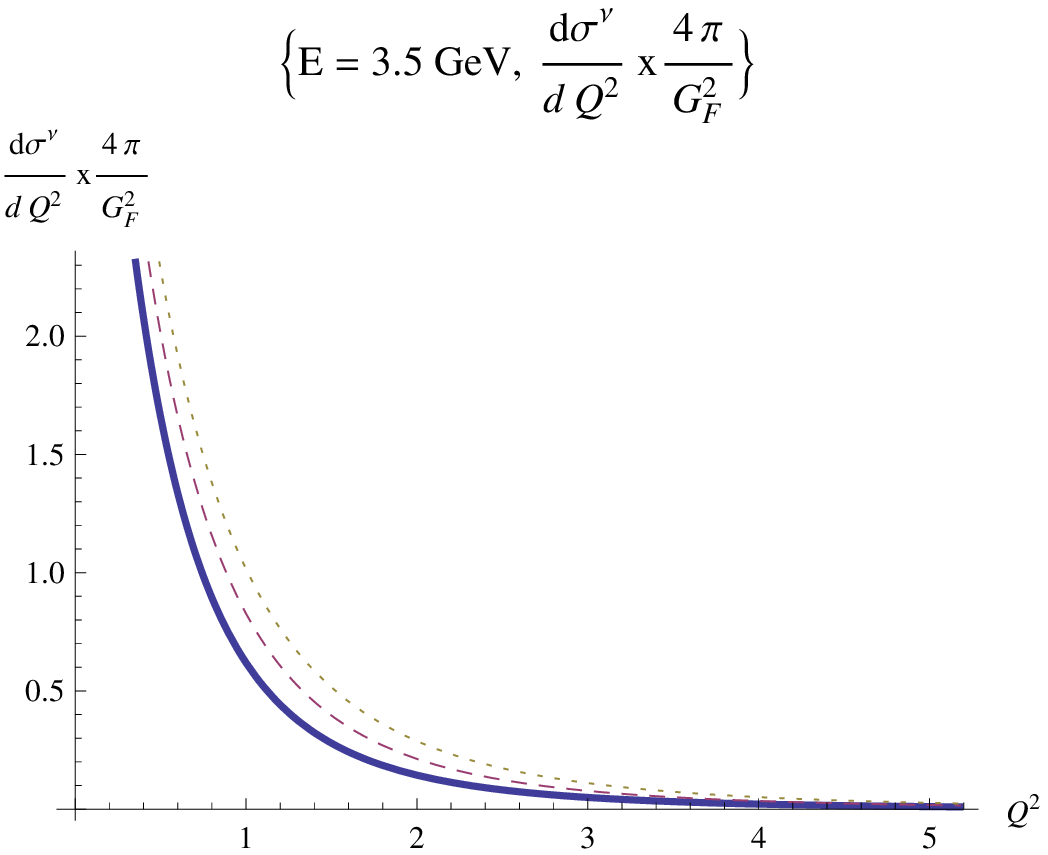,width=65mm}}

 \caption{The
dependence of the cross section (multiplied by $4\pi/G^{2}_{F}$) on the  values of $M_A$ [see(\ref{MA})] for
the processes $\bar\nu_\mu+p\to \mu^++n$  (\textit{left}) and  $\nu_\mu+n\to \mu^-+p$
 (\textit{right}) at E=1 (\textit{up}) and 3,5
 (\textit{down}) GeV.}\label{sigma}
\end{figure}

 On Figs. (\ref{sigma}) we show the differential cross sections
(multiplied by $4\pi/G^{2}_{F}$) for  the processes $\bar\nu_\mu+p\to \mu^++n$ (left)  and
$\nu_\mu+n\to \mu^-+p$ (right)
for the energies E=1 and 3.5 GeV, and  the same $M_A$, eq.(\ref{MA}). From these figures it is clear
 that it's a very difficult task  to distinguish among the different values $M_A$ solely from
 measurements of the cross sections.

\section{Conclusion}
Investigation of the CCQE neutrino processes and determination of the axial
form factor of the nucleon is of  great importance for the theory and for the modern high-precision neutrino
oscillation experiments. Many experiments on  measurement of the cross sections of the CCQE neutrino
processes in a wide range of neutrino energies have been done. From analysis of the data of these experiments
the value of the parameter $M_{A}$, which determines the $Q^{2}$-behavior of the axial form factor in the dipole approximation,
was determined. Usually in such analysis the impulse approximation for the target nuclei is used. The values of $M_{A}$ determined
from the data of the different experiments in such a way are not compatible. There could be different reasons for such a
disagreement: nuclei effects, more complicated than dipole  $Q^{2}$-dependence of the axial form factor etc.

In this paper we present the calculation of the polarization of the final nucleon in CCQE scattering.
Relations that express the axial form factor through the polarization of the final
nucleon and the electromagnetic form factors are obtained.  Our numerical analysis showed that
there is a clear sensitivity to $M_A$ in the polarizations of the neutron in $\bar\nu_\mu+p\to \mu^++n$,
most sensitive  is the ratio of the longitudinal to transverse polarization $s_\|/s_\perp$.
 This sensitivity is pronounced in the whole $Q^2$-energy range.

 We have considered the idealized case of  monochromatic neutrinos on a free nucleon.
 In order to obtain the measurable polarization the procedure of averaging of the corresponding
 expressions over the neutrino spectrum should be performed and  nuclear effects taken into account.

Experiments on  measurement of the polarization of the final proton
in elastic scattering of longitudinally polarized electrons on
unpolarized protons drastically changed our understanding about
 the electromagnetic form factors of the proton.
 Analogously, we suggest that measurement of the polarization of
the final nucleon in CCQE processes will provide additional
information about the axial form factor. It is obvious that such
measurement is a challenge. However, taking into account the
importance of the problem of the axial form factor  and the rapid
progress of the neutrino detection technique it is worth to consider
such a possibility.

\section*{Acknowledgments}

S.M. acknowledges support in part by RFBR Grant N 13-02-01442; the
work of E.Ch. is partially supported by a priority Grant between
JINR-Dubna and  Republic Bulgaria, theme 01-3-1070-2009/2013 of the
BLTP.


\end{document}